\documentclass[fleqn,9pt,onecolumn]{SelfArx} 
\definecolor{color1}{RGB}{0,0,90} 
\definecolor{color2}{RGB}{0,20,20} 
\usepackage{hyperref} 
\usepackage{epstopdf}
\usepackage{cite}
\usepackage{siunitx}
\usepackage{nicefrac}
\usepackage{fixltx2e}
\usepackage{amsmath}
\usepackage{hyphenat}
\usepackage{etoolbox}

\hypersetup{hidelinks,colorlinks,breaklinks=true,urlcolor=color2,citecolor=color1,linkcolor=color1,bookmarksopen=false,pdftitle={Title},pdfauthor={Author}}
\JournalInfo{Uploaded 6$^{\mathrm{th}}$ November 2020} 
\Archive{} 

\PaperTitle{Observing distant objects with a multimode fibre-based holographic endoscope}

\Authors{Ivo T. Leite\textsuperscript{1*}, Sergey Turtaev\textsuperscript{1}, Dirk E. Boonzajer Flaes\textsuperscript{1}, and Tom\'{a}\v{s} \v{C}i\v{z}m\'{a}r\textsuperscript{1,2,3*}}

\affiliation{\textsuperscript{1}\textit{Leibniz Institute of Photonic Technology, Albert-Einstein-Stra\ss e 9, 07745 Jena, Germany}}
\affiliation{\textsuperscript{2}\textit{Institute of Applied Optics, Friedrich Schiller University Jena, Fr\"{o}belstieg 1, 07743 Jena, Germany}}
\affiliation{\textsuperscript{3}\textit{Institute of Scientific Instruments of the Czech Academy of Sciences v.v.i., Kr\'{a}lovopolsk\'{a} 147, 612 64 Brno, Czech Republic}}

\affiliation{*\textbf{Corresponding authors}: ivo.leite@leibniz-ipht.de, tomas.cizmar@leibniz-ipht.de}

\newcommand{\keywordname}{Subject terms} 

\Abstract{\bf
Holographic wavefront manipulation enables converting hair-thin multimode optical fibres into minimally invasive lensless imaging instruments conveying much higher information densities than conventional endoscopes.
Their most prominent applications focus on accessing delicate environments, including deep brain compartments
, and recording micrometre-scale resolution images of structures in close proximity to the distal end of the instrument.
Here, we introduce an alternative `far-field' endoscope, capable of imaging macroscopic objects across a large depth of field.
The endoscope shaft with dimensions of 0.2 $\times$ 0.4 $\textrm{mm}^2$ consists of two parallel optical fibres, one for illumination and the second for signal collection.
The system is optimized for speed, power efficiency and signal quality, taking into account specific features of light transport through step-index multimode fibres.
The characteristics of imaging quality are studied at distances between 20 and 400 mm.   
As a proof-of-concept, we provide imaging inside the cavities of a sweet pepper commonly used as a phantom for biomedically relevant conditions.
Further, we test the performance on a functioning mechanical clock, thus verifying its applicability in dynamically changing environments.
With performance reaching the standard definition of video endoscopes, this work paves the way towards the exploitation of minimally-invasive holographic micro-endoscopes in clinical and diagnostics applications.   
}

\makeatletter
\newcommand{\supplabel}[2]{\protected@write \@auxout {}{\string \newlabel {#1}{#2}}}
\makeatother

\emergencystretch=\maxdimen
\begin{document}
\fontsize{3.3mm}{4.3mm}\selectfont
\flushbottom 
\onecolumn
\maketitle 


\noindent Holographic endoscopes harnessing controlled light transport through hair-thin multimode optical fibres have recently emerged as powerful technological candidates for minimally invasive observations in biomedical applications \cite{DiLeonardo2011, Cizmar2011, Cizmar2012, Papadopoulos2012, Choi2012}. 
The concept spun-out from groundbreaking research on photonics of complex media \cite{Vellekoop2007,Popoff2010b,Mosk2012,Rotter2017}, utilizing empirical quantification of optical propagation through a random medium\cite{Popoff2010}, thus heralding a new era of applications which cannot be met by conventional endoscopes.  
Their recognized potential to obtain detailed imagery from large depths of sensitive tissue structures \cite{Frank2019} has recently been exploited in {\it in-vivo} neuroscience studies \cite{Ohayon2018,Turtaev2018,Vasquez-Lopez2018}, where neurones and processes of neuronal circuits have been acquired in living animal models through fibres having footprints of $\approx$\SI{0.01}{\milli\metre\squared}.

\begin{figure*}[t] 
   \centering
   \captionsetup{justification=justified}
   \includegraphics[width=\textwidth]{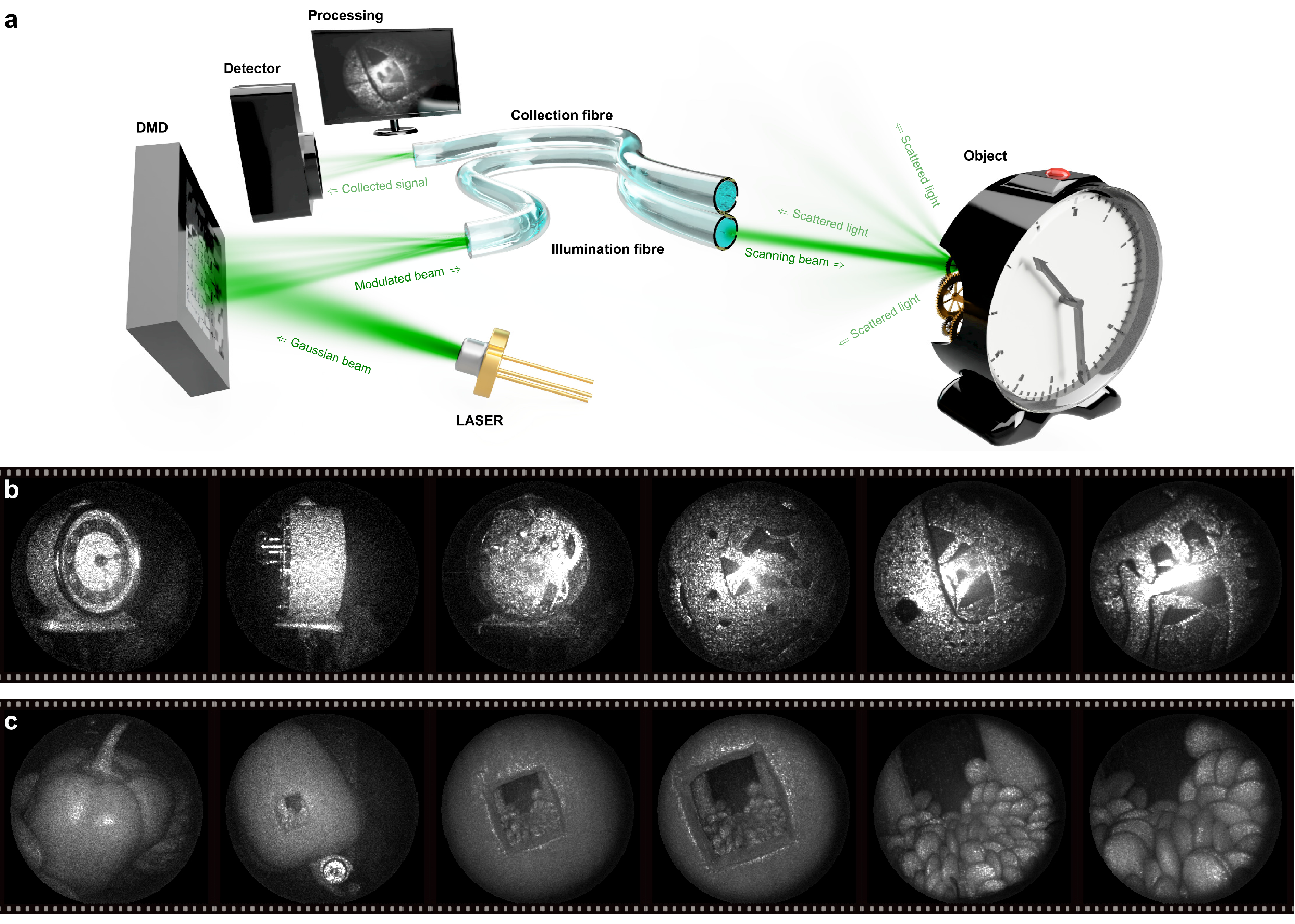}
   \caption{ \textbf{Multimode fibre based far-field endoscope.}
   		\textbf{a,} A sequence of holograms displayed by a digital micro-mirror device (DMD) spatially shapes the wavefronts coupled into a multimode optical fibre, in such a way that a far-field focus scans the distal field of view. The light signals back-scattered by an object are partially captured by a collection fibre, allowing real-time image reconstruction at the proximal side.
   		\textbf{b,}  Imaging a mechanical clock as an example of a dynamic object. 
   		\textbf{c,} Demonstration of endoscopic imaging of a sweet pepper via a small opening. 
   		Full video recordings are available as Supplementary Materials.
   	\label{fig:concept}}
\end{figure*}
The aspiration of this work is to extend the applicability of holographic endoscopes into clinical environments, where reduction of instrument's footprint is equally desired, yet the demands on imaging performance differ considerably from those of \textit{in-vivo} neuroscience.
This relates not only to spatial resolution and frame rate, but importantly to working distance and field of view, which have to be significantly enhanced before the concept can be accepted as a credible strategy for new minimally-invasive diagnostics and surgery-assisting instrumentation.  
In this work we therefore focus on imaging objects placed at macroscopic distances away from the instrument's distal end, and address the limitations which arose from this new regime. 
Reaching high imaging speed dictates short pixel dwell times, nowadays only possible with power-inefficient devices \cite{Turtaev2017}.
Moreover, imaging distant objects combined with further minimizing the instrument's footprint dramatically reduces the amount of photons returning from the object through the endoscope shaft.
Therefore the overall power efficiency and wavefront-control fidelity represent a major challenge.
While initial studies in this field have treated multimode fibres as entirely random media, it has been shown recently that light transport at the distances relevant for endoscopy is almost perfectly predictable in straight as well as significantly deformed fibres \cite{Ploschner2015a}.
Even utilizing a fragmental knowledge about the optical transmission through various input-output correlations may lead to startling benefits in practical operation of fibre-based endoscopes \cite{Li2020}.
Amongst other technological advancements described below, the potential of light-transport predictability is exploited in this work by designing the system in a manner benefiting from radial \textit{k}-space conservation observed in step-index multimode fibres.
In the far-field imaging regime developed here, this alone greatly enhances the power efficiency and associated signal-to-noise ratio of the resulting imagery. 
To provide a qualitative appreciation of the endoscope system, we showcase the imaging performance inside a sweet pepper and a functioning clockwork mechanism.
Further, we provide a detailed quantitative analysis of the imaging fidelity for various distances, and compare the radial \textit{k}-space conservation benefiting system with the commonly used geometry.
Although compressive sensing or machine-learning algorithms could accelerate imaging or eliminate inherently speckled nature of the images \cite{Amitonova2018,Rahmani2018,Borhani2018}, all our results are presented as raw measurements with their original contrast in order to avoid any biases.

\section*{Results}
\subsection*{The far-field endoscope in action}
The endoscope exploits the principle of raster-scan imaging, whereby images are reconstructed from the local response of an object to a scanning beam  pre-shaped by the holographic modulator and delivered by a multimode optical fibre.
In its simplest implementation, the endoscope collects photons back-scattered by the object, with their amount depending on the object's local reflectivity, roughness, orientation, and axial depth.
Because only a small portion of the back-scattered light falls within the small collection area defined by the fibre core, at large imaging distances the detected signals become feeble in comparison to the constant background formed by reflections at the air/glass interfaces of the fibre and other optical components in the system.
For this reason, we collect the reflected light signals by a separate fibre, which guides them towards a highly-sensitive bucket detector, as illustrated in Fig.~\ref{fig:concept}a.
A fast digital micro-mirror device (DMD) is employed as the wavefront shaping element in the system.
Light propagation through the illumination fibre is empirically characterised by a transmission matrix describing the linear relationship between conveniently chosen sets of input and output fields. 
In our case, we use the representation of orthogonal plane-waves truncated by the DMD chip as the basis of input fields, and diffraction-limited foci across a square grid in the far-field plane of the distal fibre facet as the basis of output fields.
Once acquired, the transmission matrix contains the information for designing the binary DMD patterns for pre-shaping the proximal wavefront which result in far-field foci on the distal end of the endoscope.
A detailed description of the setup and holographic methods is given in Supplementary Methods.

Imaging of three-dimensional, natural scenes is shown in Fig.~\ref{fig:concept}b and c.
The field of view is scanned by approximately \num{100000} far-field foci, yielding greyscale images with \num{0.1}~megapixel definition, a value generally accepted as the standard definition of video endoscopes \cite{Bhat2014}.
Figure~\ref{fig:concept}c shows selected frames from a video recording of a yellow sweet pepper, an object commonly used for testing endoscope performance since its cavities resemble orifices of higher-order organisms.
A small opening carved on the lateral surface facilitated the access of the endoscope to observe its interior.
The full video recording is available as Supplementary Movie SM1.
Being a far-field imaging system, the field of view scales linearly with the imaging distance, provided the far-field condition is fulfilled.
This can be understood in terms of the Fresnel number which, for the illumination fibre used (\SI{200}{\micro\metre}-diameter core) and \SI{532}{\nano\metre} wavelength, translates into imaging distances greater than \SI{2}{\centi\metre}.
Naturally, a trade-off exists between the accessible field of view and the attainable spatial resolution, resulting from the fixed angular resolution of the endoscope.
This is addressed in greater depth in Supplementary Results~1.

As in any coherent imaging technique, the reconstructed images are affected by speckle artefacts.
A spatially coherent far-field focus, back-scattered by a diffusive object, gives rise to speckle with grain size exceeding that of the collection area of the endoscope.
As the sample is scanned by identical far-field foci, the backscattered speckle field changes accordingly and overlaps differently with the collection core, manifesting itself as fluctuations in the detected intensity.

The imaging pixel rate is limited by the \SI{23}{\kilo\hertz} refresh rate of the state-of-the-art spatial light modulator employed, resulting in a \SI{4.4}{\second} acquisition time for each \num{100}~kilopixel image shown here.
When imaging a dynamic scene such as the rotating clockwork mechanism shown in Fig.~\ref{fig:concept}b, this can lead to motion artefacts.
To mitigate this effect, the spatial sampling was decreased on-the-fly to speed up the acquisition frame rate to 0.91~fps, 3.6~fps, and 15~fps -- resulting in 25-, 6.3-, and 1.6-kilopixel images, respectively -- as shown in Supplementary Movie SM2.
Random access to the sequence of holograms loaded to the DMD allows also restricting raster-scanning to a particular region of interest, which can be imaged at a frame rate scaling inversely with its number of pixels.
This is discussed in greater detail in Supplementary Results~2.

\subsection*{Focusing in the far field}

\begin{figure*}[ht] 
	\centering
	\captionsetup{justification=justified}
	\includegraphics[width=\textwidth]{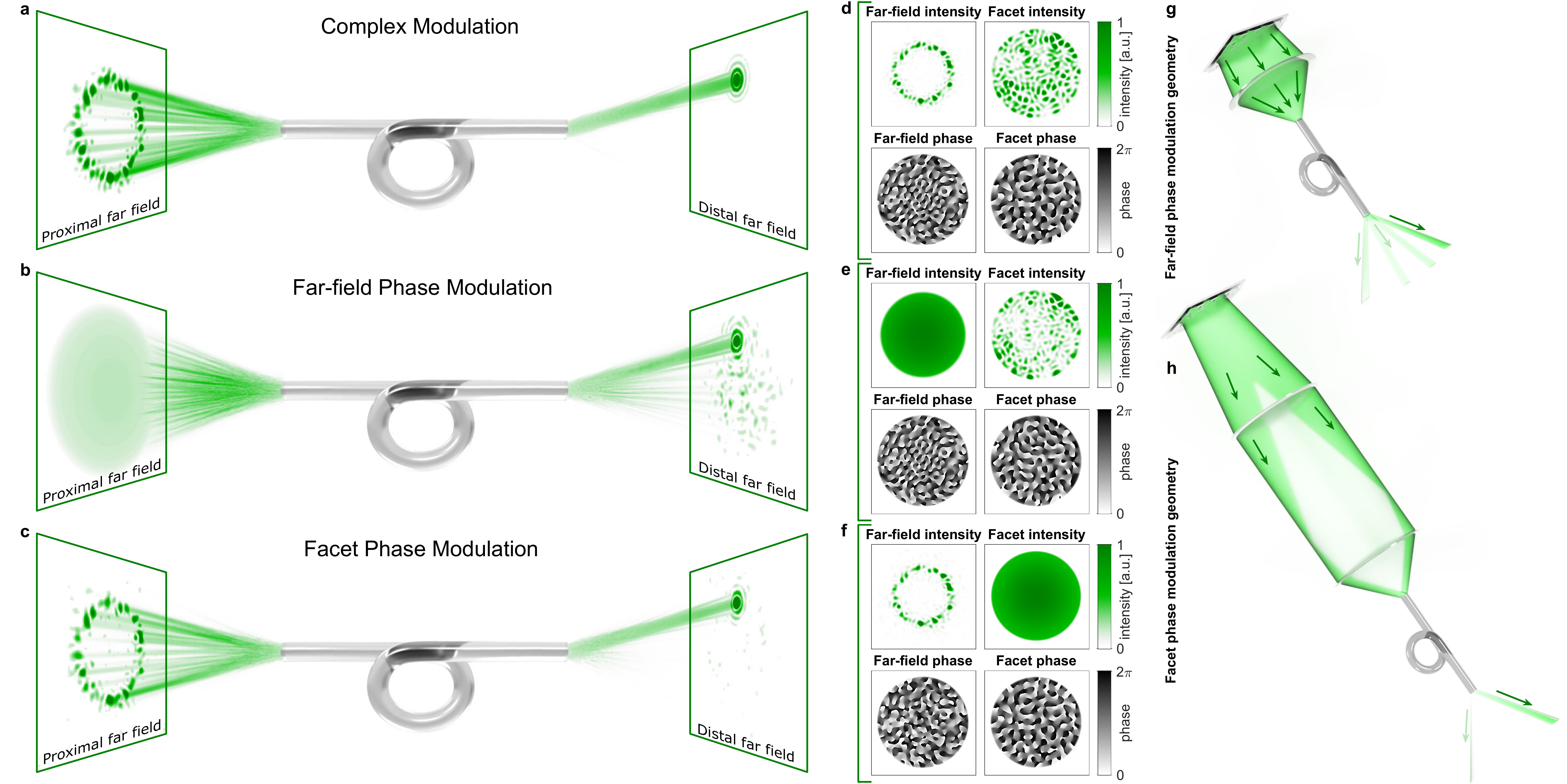} 
	\caption{\textbf{Far-field focusing through a step-index multimode fibre.}
		\textbf{a,} Due to the conservation of propagation constants in step-index fibres, only light signals originating from a particular annular region in the proximal far field can contribute to synthesize a distal far-field focus at a given distance from the optical axis.
		In the ideal case when complex modulation with arbitrary resolution is applied, all incident power is found in the target far-field focus.
		\textbf{b,} In case of far-field phase modulation, the input spectrum contains all spatial frequencies supported by the fibre, and a background signal arising from uncontrolled light is found on the distal far field spanning its entire acceptance cone.
		\textbf{c,} When phase modulation takes place in a plane conjugate to the proximal fibre facet, the optical power can be efficiently redistributed in the proximal far field such that it closely matches the ideal target field, thus increasing the fraction of controlled power.
		\textbf{d-f,} Intensity and phase profiles of the proximal fields coupled into the fibre for the modulation approaches depicted in (a-c).
		\textbf{g,h,} Practical implementation of phase-only modulation in the far field (\textbf{g}) and in a plane conjugate to the proximal fibre facet (\textbf{h}).
	\label{fig:design}}
\end{figure*}

Despite its apparently randomizing nature, light transport through multimode optical fibres has properties differing from other complex media.
All multimode fibres support a number of propagation invariant modes (PIMs) which do not change their field distributions as they propagate through the fibre, each of which being characterized by a particular propagation constant dictating its phase velocity \cite{Snyder1983}.
Within distances relevant for endoscopic applications, energy exchange between PIMs is typically negligible \cite{Ploschner2015a}.
Specifically for step-index fibres, which confine most of the optical power in their optically homogeneous core, PIMs entering or leaving the fibre from and to other homogeneous media are quasi-non-diffracting beams also featuring extremely narrow distributions of propagation constants.
Their far-field intensities therefore take the shape of an annulus with diameter increasing with decreasing propagation constant of the PIM.
This leads to an input-output correlation between the proximal and distal far fields of a step-index multimode fibre, where power from a certain annular zone in the proximal far field is delivered to the corresponding annular region in the distal far field without mixing with others.
This has been efficiently utilized for refocusing diffraction-limited foci \cite{Cizmar2012,Leite2018}, as well as synthesizing Bessel beams in the close proximity of the fibre facet \cite{Cizmar2012,Ploschner2015b}.
In relevance to far-field endoscopy, creating a focus at the distal far field is equivalent to producing at the distal fibre facet a plane-wave truncated by the fibre core, i.e. an optical field also featuring an extremely narrow distribution of propagation constants.
The corresponding proximal far field is therefore a superposition of PIMs with very similar annular far-field intensity distributions.
They feature diverse topological charges (azimuthal phase dependencies) and mutually intricate phase relationships due to the macroscopic lengths of fibre, and thus their coherent sum leads to an apparently random character of speckles.
Yet, the optical power remains concentrated in the commonly shared annular zone, as illustrated in Fig.~\ref{fig:design}a.
This simulation represents the ideal case where both the phase and amplitude of the incident wavefront are modulated with arbitrary resolution, so that all optical power contributes to form a desired far-field focus. 
However, imprinting complex modulations onto a wavefront generally incurs prohibitive power losses, as it involves discarding most of the available power, therefore in practice, phase-only modulation is usually employed. 
Since holographic modulators have gradually replaced beam-steering and wavefront-correcting elements, traditional wavefront-shaping systems adopt a far-field phase modulation arrangement, as depicted in Fig.~\ref{fig:design}b, where the incident fields are pre-shaped by a spatial light modulator at the back-focal plane of the focusing optics -- i.e. the Fourier plane of the proximal fibre facet -- as illustrated in Fig.~\ref{fig:design}g (designs frequently use relay optics omitted here).  
In this case, light signals originating outside the aforementioned annular region in the proximal far field cannot contribute to the target far-field focus, instead giving rise to a speckle background spanning the acceptance cone of the fibre, and thus decrease the fraction of controlled power. 

The alternative arrangement exploited here performs the phase modulation in the facet plane, allowing an efficient redistribution of the available power in the proximal far field, as shown by the simulation in Fig.~\ref{fig:design}c.
This implementation involves placing the wavefront shaping element in a plane conjugate to the proximal fibre facet, as illustrated in Fig.~\ref{fig:design}h, and allows generating a modulated field closely resembling the target ring-shaped field in the proximal far field, from phase-only modulations.
As a consequence, only the required spatial frequencies are coupled into the fibre, and just a small fraction of stray power ($\approx$ 20\%) reaches the sample plane in this regime \cite{Chandrasekaran2014}. 

In the following, we compare the performance of the endoscope system in the two competing configurations discussed. For simplicity, these geometries are referred to hereinafter as ``Fourier plane'' and ``Field plane'', indicating the mutual placement of the DMD and the proximal fibre facet.
Phase-only modulation is achieved from binary-amplitude DMD patterns using the Lee hologram approach \cite{Lee1979,Conkey2012,Turtaev2017}.


\subsection*{Imaging performance analysis}

\begin{figure*}[h] 
   \centering
   \captionsetup{justification=justified}
   \includegraphics[width=\textwidth]{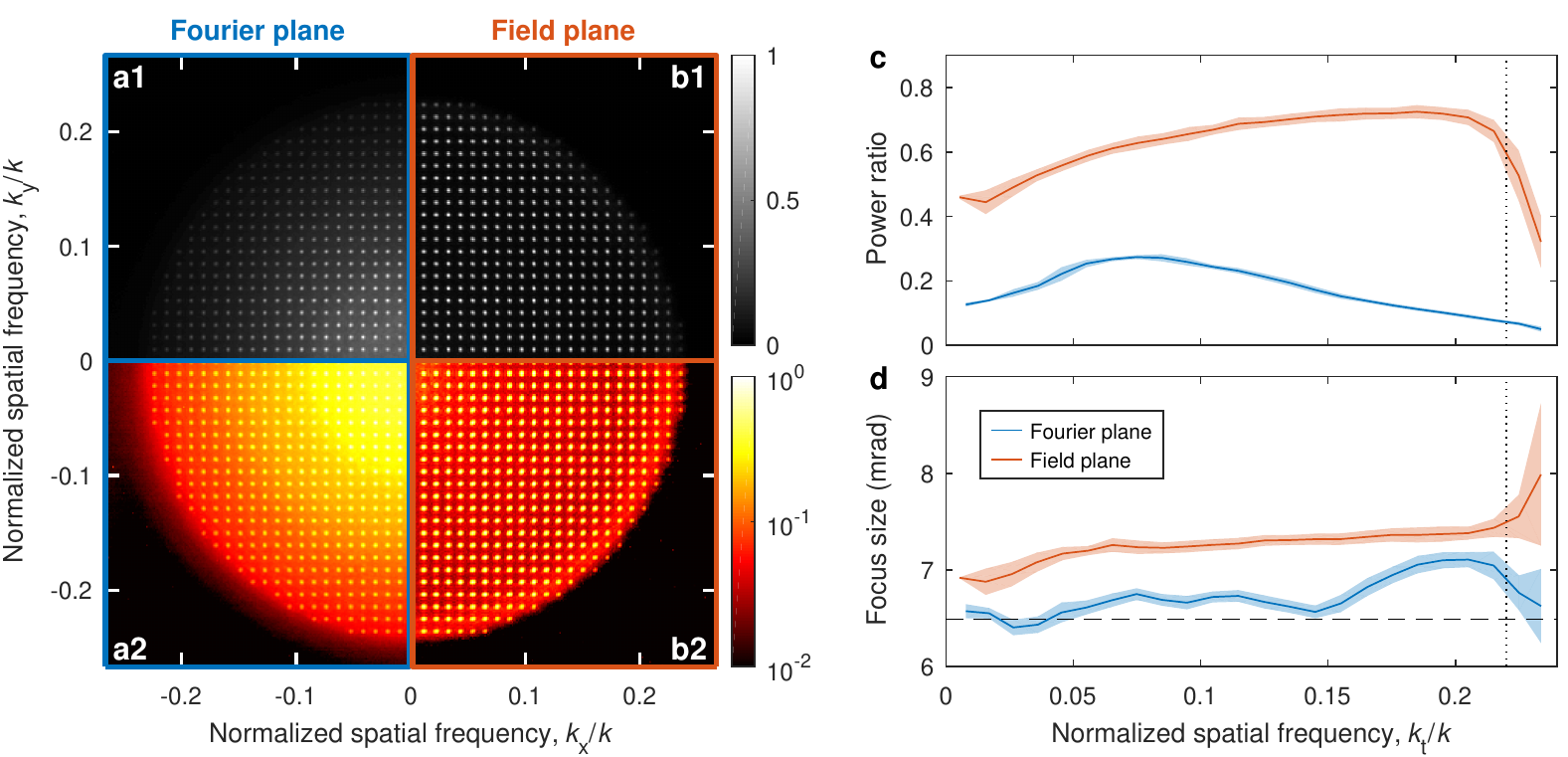}
   \caption{\textbf{Focusing quality in the far field.}
   		\textbf{a,b,} Sum projection of several far-field foci generated along a square grid in the ``Fourier plane'' (\textbf{a}) and in the ``Field plane'' (\textbf{b}) configurations, shown both in linear (a1,b1) and logarithmic (a2,b2) scales.
   		\textbf{c,d,} Comparison of the focusing quality of the two designs in terms of the fraction of power contained in the desired focus (\textbf{c}) and of its angular size (\textbf{d}), as function of the normalized spatial frequency (i.e. distance to the optical axis in the far field) expressed as $\nicefrac{k_t}{k}$, where $k_t^2=k_x^2+k_y^2$ and $k$ is the wavenumber in free space.
   		The horizontal dashed line in (d) indicates the theoretical limit of \SI{6.5}{\milli\radian}.
   		In (c,d), the vertical dotted line indicates the nominal NA of the fibre, \num{0.22}, and the shadowed areas indicate standard deviations around the mean values.
	\label{fig:foci}}
\end{figure*}

Figure~\ref{fig:foci} shows a comparison of the foci quality generated in the competing geometries introduced above.
Panels (a) and (b) show the sum projection of approximately \num{1500} far-field foci generated sequentially across a sparse square grid limited by the numerical aperture of the illumination fibre.
The power uniformity is visibly better in the ``Field plane'' configuration (b), and does not vanish with increasing angles (expressed in terms of normalized spatial frequency) until the edge of the field of view is reached.
A much smaller background level can also be seen in this geometry, indicating that a smaller fraction of the output power falls outside the foci.
The fraction of transmitted optical power contributing to the desired light output, termed \textit{power ratio}, is used here as  a metric to quantify the fidelity of the generated foci -- as discussed in greater detail in Supplementary Results~3 -- since it directly impacts on the contrast of the resulting imagery.

The intensity distribution of each such far-field focus was fitted to an Airy pattern.
This allowed estimating both their power ratio, retrieved as the fraction of power under the fitted surfaces with respect to the total power emerging from the fibre, as well as their size, taken as the angular aperture of the first-zero ring in the Airy distributions.
Figures~\ref{fig:foci}c and d show the power ratio and size of the generated foci as a function of their radial coordinate in the far field, for both geometries under test.
The fraction of controlled power is shown to be improved when the proximal fibre facet is placed in the field plane of the DMD, confirming the hypothesis that a larger portion of the input power gives rise to background signal when it is placed in the Fourier plane of the spatial light modulator.
The ``Fourier plane'' configuration is able to generate far-field foci closely matching the theoretical value of \SI{6.5}{\milli\radian} expected for an Airy disk resulting from diffraction of a plane-wave truncated by the \SI{200}{\micro\metre}-diameter fibre core, at the used wavelength.
The ``Field plane'' geometry, however, produces around 8\% larger foci, which we believe is caused by the Gaussian envelope of the signal illuminating the DMD chip which, due to the nature of light transport described above, affects the power distribution at the distal facet. As shown below, this leads to a relatively small reduction of imaging resolution, which can be improved by employing a Gauss-to-top-hat beam-shaping element in front of the DMD.

\begin{figure*}[h] 
   \centering
   \captionsetup{justification=justified}
   \includegraphics[width=\textwidth]{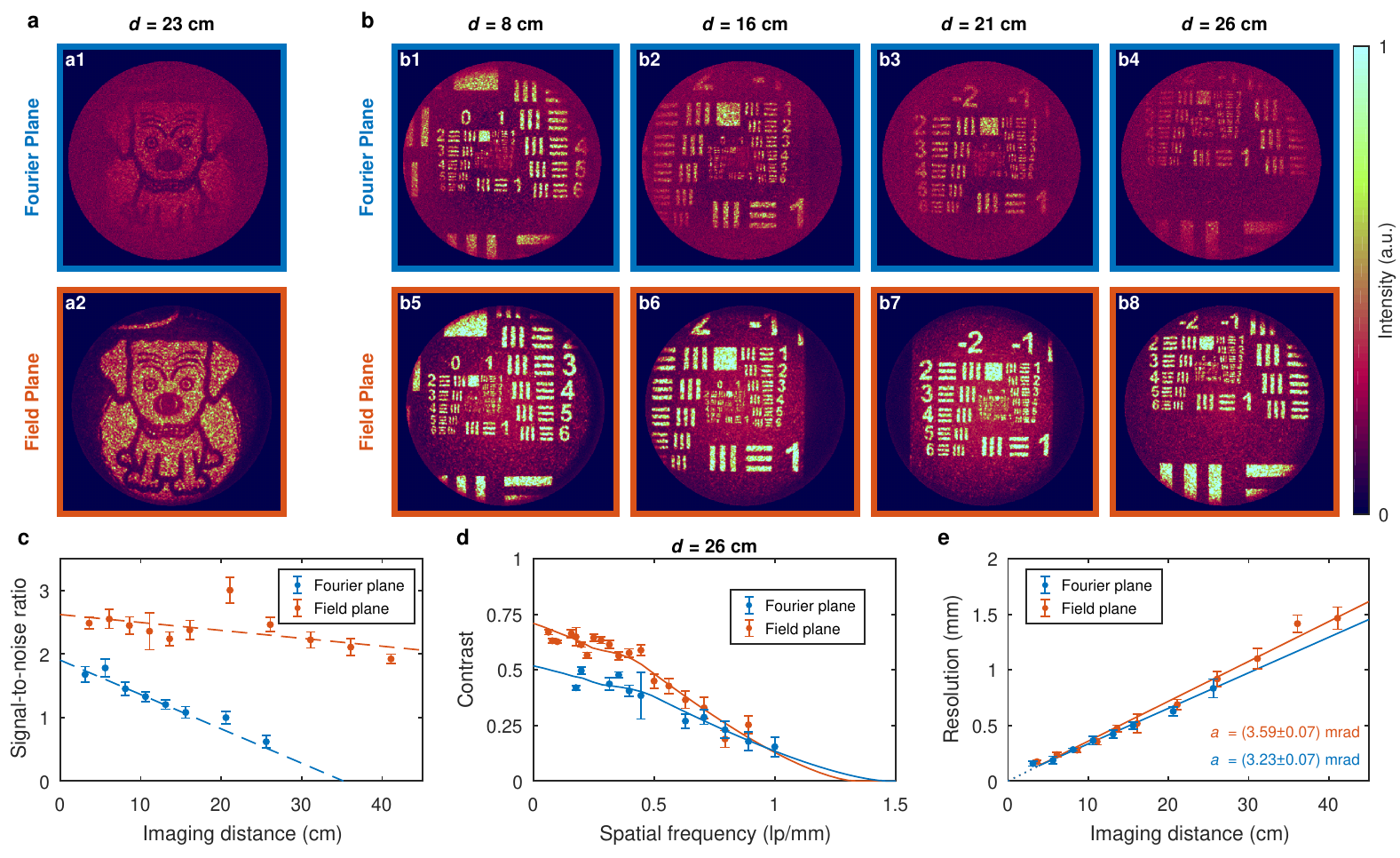}
   \caption{\textbf{Assessment of imaging performance.}
   		{\bf a,} Imaging a two-dimensional, Lambertian object under each design geometry.
   		{\bf b,} Imaging a negative 1951 USAF resolution test chart placed at varying distances $d$ from the endoscope termination.
   		{\bf c,} Signal-to-noise ratio retrieved from the test chart images, as function of imaging distance.
   		{\bf d,} Contrast of several test chart elements as function of their spatial frequency (expressed in line pairs per millimetre, lp/mm) for an imaging distance of \SI{26}{\centi\metre}, fitted to the contrast transfer function (CTF).
   		{\bf e,} Imaging resolution retrieved from the CTFs, as function of imaging distance.
   		The slopes in (e) correspond to the angular resolution of the system in each configuration.
   		The errorbars in (c-e) indicate standard deviations.    
   \label{fig:imaging}}
\end{figure*}

An image-based assessment of performance is given in Fig.~\ref{fig:imaging}, where each image in Fig.~\ref{fig:imaging}a and b is normalized to 5 times its median intensity. 
An imaging comparison between the two configurations is shown in Fig.~\ref{fig:imaging}a for a two-dimensional, quasi-Lambertian object, placed at approximately \SI{23}{\centi\metre} from the endoscope.
Higher image contrast is clearly obtained in the ``Field plane'' configuration, especially when approaching the periphery of the field of view where the power ratio becomes very small for the ``Fourier plane'' geometry.
A negative 1951 USAF resolution test chart (comprising groups -4 to 7) was imaged at a varying distance from the fibre endoscope, with the output optical power limited to approximately \SI{0.1}{\milli\watt}.
Figure~\ref{fig:imaging}b shows selected examples for four imaging distances in each design configuration.
In addition to their aforementioned speckled nature, the images are also affected by detection noise, particularly for larger imaging distances when very few back-scattered photons are effectively captured by the small collection area of the endoscope.
This is quantified in Fig.~\ref{fig:imaging}c, which shows the signal-to-noise ratio of the reconstructed images as a function of imaging distance.
We should note that the gain of the photomultiplier tube detector used as well as the output optical power were re-adjusted at each imaging distance, as described in the Methods section, and that the sparsity of the images depends on the particular region of the object being examined. 
The ``Field plane'' geometry yields higher signal fidelity, again due to the increased power ratio achieved in this configuration.

The images obtained of the 1951 USAF test chart were used for further assessing the imaging performance.
At any given imaging distance, the contrast of the target elements as function of their spatial frequency follows the contrast transfer function (CTF), which describes the response of the imaging system to a square wave (a binary black and white stripe pattern).
This is exemplified in Fig.~\ref{fig:imaging}d, where the fitted CTF was computed using Coltman's formula \cite{Coltman1954}, from the modulation transfer function of a diffraction-limited imaging system with a circular aperture. The fitting procedure searched for two free parameters, corresponding to the amplitude (i.e. zero-frequency contrast) and cut-off spatial frequency.
The latter allows retrieving the spatial resolution (Rayleigh's definition), which is plotted against imaging distance in Fig.~\ref{fig:imaging}e.
Here, the slopes of the fitted lines correspond to the angular resolution of the system, which is found to be \SI{3.59(7)}{\milli\radian} and \SI{3.23(7)}{\milli\radian} for the ``Field plane'' and ``Fourier plane'' configurations, respectively.
The imaging resolution is slightly higher when the input fibre facet is placed in the Fourier plane of the DMD, rather than in the field plane, which is consistent with the observation that the far-field foci are smaller in size in this configuration, shown above in Fig.~\ref{fig:foci}d.

While both the traditional ``Fourier plane'' geometry and the alternative ``Field plane'' configuration exploited here lead to far-field images with comparable spatial resolution, the latter is shown to yield images with far improved signal-to-noise ratio owing to the increased degree of control over the output light fields forming diffraction-limited foci in the distal far field.

\section*{Discussion}

To the best of our knowledge, in this work we show the first far-field endoscope based on multimode fibre.
Due to its extended depth of field, the holographic endoscope is capable of imaging macroscopic objects in a wide range of distances.
Here we reach the current technological limits, employing the fastest available spatial light modulator and having an unprecedented control over \num{17000} fibre modes.
This allowed transmitting \num{0.1}~megapixel images, thus reaching the standard definition of modern video endoscopes \cite{Bhat2014}.
The versatility of the instrument was demonstrated by imaging complex, three-dimensional scenes, particularly the interior of a sweet pepper serving as a phantom for biomedically relevant environments, as well as a functioning clockwork mechanism as an example of an object with dynamic complexity.
Because of radial \textit{k}-space conservation of propagating fields in step-index fibres, a larger fraction of the optical power can be directed towards a far-field focus when phase-only modulation is performed in the field plane of the input fibre facet, rather than in its Fourier plane.
This allows efficiently manipulating the spatial frequency spectra of the coupled fields, and results in far-field imaging with greatly improved signal-to-noise ratio.

The field of view, resolution and minimum imaging distance of the endoscope depend solely on intrinsic parameters of the illumination fibre, namely its numerical aperture and core size, whereas the number of resolvable image features (which impact on the definition of the resulting imagery) depend on the number of guided modes it supports.
As such, the imaging properties of the system can be tailored to the particular needs of the envisioned application by selecting the suitable fibre, as discussed to a greater extent in Supplementary Results~2. 
With the numerical aperture of currently available fibres now reaching up to \num{0.9} \cite{Leite2018}, identical \num{0.1}~megapixel images could be obtained using a \SI{50}{\micro\metre}-diameter core illumination fibre, with the instrument footprint decreased by 16 fold.
Naturally, the probe size could be further halved by adopting a single fibre design.
A single core fibre can serve the dual purpose of illuminating the sample and collecting back-scattered photons, provided the reflections at the fibre endfaces are eliminated -- using anti-reflective coatings or angled terminations -- or by exploiting alternative strategies for discriminating between the light signals, such as separating them in the time domain.
The endoscope can also be extended to colour imaging by calibrating the illumination fibre at multiple wavelengths.
The different illumination colours can be either scanned sequentially, or even simultaneously by combining them into optimised holograms, with the back-scattered signals being separated spectrally on the proximal side.
Finally, the speckled nature of the reconstructed images can in principle be suppressed computationally using a multitude of available strategies \cite{Bianco2018}.

This work sets the scene for the introduction of semi-rigid, minimally-invasive multimode fibre probes as a perspective alternative to the rigid endoscopes routinely used in clinical diagnostics and key-hole surgery.
Relying in its heart on a reconfigurable device, this endoscope is compatible with a multi-modal operation, whereby dynamically switching from far-field imaging to detailed microscopic observations near the fibre is possible simply by updating the holograms displayed by the spatial light modulator.
Both hologram sequences can be obtained from two transmission matrices acquired simultaneously, or numerically by multiplying either one with a free-space propagation operator \cite{Ploschner2015a}.
The far-field modality can thus be used to guide the insertion of the probe while providing organ- or tissue-scale imagery, which can then be promptly converted into an instrument imaging at the subcellular scale.
Combined with spectroscopic imaging methods \cite{Gusachenko2017,Deng2019,Tragardh2019}, the technology has potential for \textit{in situ} diagnostics at the cellular level.

\bibliographystyle{naturemag_noURL}
\bibliography{PepperRefs.bib}

\section*{Acknowledgements}
The authors acknowledge the support from the European Research Council (724530), the European Regional Development Fund (CZ.02.1.01/0.0/15\_003/0000476), the Th\"{u}ringer Ministerium f\"{u}r Wirtschaft, Wissenschaft und Digitale Gesellschaft, the Th\"{u}ringer Aufbaubank, and the Federal Ministry of Education and Research, Germany (BMBF).
Martin \v{S}iler is gratefully acknowledged for a C library allowing efficient parallel computation of the DMD patterns.

\section*{Author contributions}
I.T.L., S.T. and T.\v{C}. performed all experiments.
I.T.L. and D.E.B.F. analyzed the results.
T.\v{C}. conceived and led the project.
I.T.L., S.T. and T.\v{C}. wrote the manuscript.

\section*{Materials \& Correspondence}
Correspondence and requests for materials should be addressed to I.T.L. or T.\v{C}.

\section*{Competing interests}
The authors declare no competing interests.

\section*{Methods}

\subsection*{Foci analysis}
Because the intensity distributions of the far-field foci (including the speckled background) shown in Fig.~\ref{fig:foci} span over the 8-bit depth of the CMOS camera used, several image frames were acquired for each focus with increasing exposure times until the speckled background was visible.
This allowed reconstructing high dynamic range images $I\left(x,y\right)$ of the intensity distributions for every far-field focus. 
Each of these was fitted to an Airy pattern:
\begin{equation}
f\left(x,y\right) = A\left(\frac{2J_1\left(\nicefrac{\rho}{w}\right)}{\nicefrac{\rho}{w}}\right)^2+c,
\end{equation}
where $J_1$ is the Bessel function of the first kind of order one, $A$ the amplitude, $c$ the offset, \(\rho=\sqrt{\left(x-x_0\right)^2+\left(y-y_0\right)^2}\) the radial polar coordinate centred around $\left(x_0,y_0\right)$, and $w$ measures the width of distribution.
The radii of the Airy disks were calculated from the fitted surfaces as $\approx 3.8317w$ (the factor being the first root of $J_1$), whereas the power ratio, $\eta$, of each focus was estimated from the fit parameters as:
\begin{equation}
\eta = \frac{\iint{\left[f\left(x,y\right)-c\right]dxdy}}{\iint{I\left(x,y\right)dxdy}}
= \frac{\nicefrac{4\pi A}{w^2}}{\sum_{x,y}{I\left(x,y\right)}}.
\end{equation}

\subsection*{Imaging methods}
Upon the calibration procedure, where the transmission matrix of the fibre is measured and the sequence of holograms for generating each far-focus is calculated, the calibration module was removed and replaced by the object to be imaged.
The sequence of binary holograms was arranged in such way as to perform 2D interlaced scanning of the object, where for each image frame acquired the field of view is scanned by successive foci generated along sparse square grids of increasing spatial frequency (i.e. decreasing spacing between foci positions), without repeating the same foci positions.
This allowed modifying the spatial sampling on-the-fly by displaying the hologram sequence only partially, thus providing a convenient means of adjusting the image acquisition frame rate.

When imaging objects at the smaller distances, the gain of the photomultiplier tube (PMT) detector was set to its minimum value, and the optical power adjusted to obtain the maximum possible output voltage signals from the detector.
As imaging distances increased, the optical power was also increased to the maximum values permitted by the PMT.
Once the maximum available output optical power was reached ($\approx\SI{0.1}{\milli\watt}$), the PMT gain was also adjusted to maximize the output voltage signals from the PMT detector.

\subsection*{Image analysis}
\subsubsection*{Signal-to-noise ratio}
Each recorded image $I\left(x,y\right)$ of the 1951 USAF test chart was compared to a binary image $B\left(x,y\right)$, shifted and re-scaled to best fit $I\left(x,y\right)$, serving as a ``ground truth''.
The signal-to-noise ratio (SNR) of each image $I$ was then retrieved as:
\begin{equation}
\textrm{SNR} = \frac{\frac{1}{N_1}\sum_{x,y}{I.B} - \frac{1}{N_0}\sum_{x,y}{I\left(1-B\right)}}{\frac{1}{N_0}\sum_{x,y}{I\left(1-B\right)}},
\end{equation}
where $N_1 = \sum_{x,y}{B}$ and $N_0 = \sum_{x,y}{\left(1-B\right)}$ are respectively the number of ``ones'' and ``zeros'' in each binary image $B$. 

\subsubsection*{Spatial resolution}
For every image of the 1951 USAF test chart, each individual target element was fitted to a 2D surface representing three parallel bars to retrieve their width.
For elements with vertical bars (i.e. oriented in the $y$ direction) the fitted surface had the form: 
\begin{equation}
f\left(x,y\right) = \sum_{m=0}^{5}{\left(-1\right)^m\frac{a}{2}\textrm{erf}\left(b\left[x-x_0-mw\right]\right)} + c,
\end{equation}
where $\textrm{erf}\left(\right)$ is the error function, $w$ is the width (and separation) between bars, $a$ is an amplitude, $b$ a measure of the ``steepness'' of the transition, $c$ an offset, and $x_0$ the horizontal position of the first bar.
The contrast of each target element was estimated as $\nicefrac{\left(f_\textrm{max}-f_\textrm{min}\right)}{\left(f_\textrm{max}+f_\textrm{min}\right)}$, where $f_\textrm{max}$ and $f_\textrm{min}$ are respectively the maximum and minimum of the fitted surface $f$.

The contrast transfer function (CTF) was computed from the modulation transfer function (MTF) using the first 100 terms of Coltman's formula:
\begin{equation}
\textrm{CTF}\left(\nu\right) = \frac{4}{\pi}\sum_{n=0}^{\infty}{\frac{\left(-1\right)^n}{2n+1}\textrm{MTF}\left(\left[2n+1\right]\nu\right)},
\end{equation}
where MTF is the modulation transfer function and $\nu$ the spatial frequency.
We consider the case of an ideal imaging system with a circular aperture, with MTF given by:
\begin{equation}
\textrm{MTF}\left(\nu\right) = A\left|\frac{2}{\pi}\left(\arccos{\left(\left|\nu^{\prime}\right|\right)}-\left|\nu^{\prime}\right|\sqrt{1-\nu^{\prime 2}}\right)\right|,
\end{equation}
where $\nu^{\prime}=\nicefrac{\nu}{2\nu_0}$ is the spatial frequency normalized to the cut-off value, and $A$ is an amplitude.

\end{document}


\fontsize{3.3mm}{4.3mm}\selectfont
\flushbottom 
\twocolumn


\renewcommand\thefigure{S\arabic{figure}}
\renewcommand\theequation{S\arabic{equation}}
\setcounter{figure}{0}
\setcounter{page}{1}
\setcounter{equation}{0}

\section*{\Large Supplementary Methods}

\begin{figure*}[h!] 
	\centering
	\captionsetup{justification=justified}
	\includegraphics[width=\textwidth]{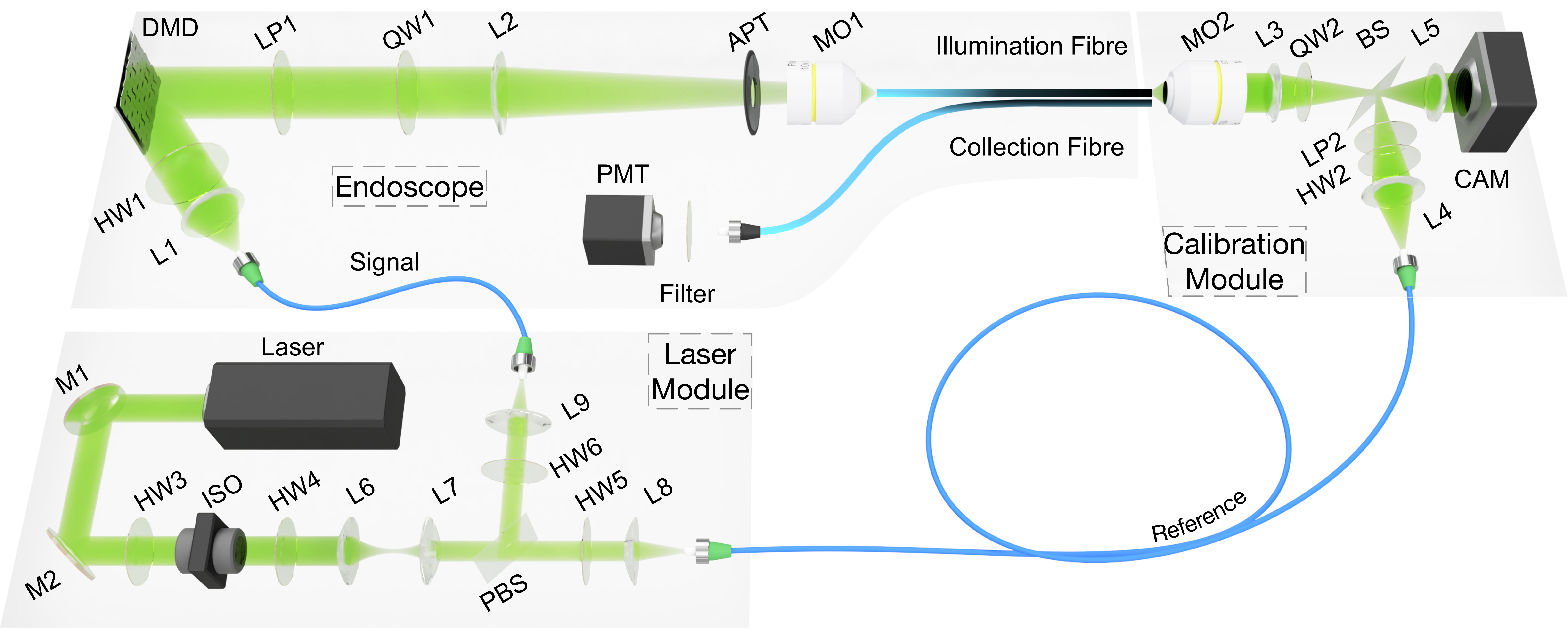}
	\caption{\textbf{Detailed setup layout of the far-field holographic endoscope.}
		Schematic representation of the system at the calibration stage, in the ``Field plane'' configuration where the DMD is imaged onto the proximal facet of the illumination fibre.
		Upon measurement of its transmission matrix, the calibration module is removed and the endoscope is ready for imaging.    
		\label{fig:setup}}
\end{figure*}

\section{\normalsize Detailed description of the system}
The optical system is depicted schematically in Fig.~\ref{fig:setup}, in the configuration where the digital micromirror device (DMD) is imaged onto the proximal endface of the illumination fibre, i.e. the geometry referred to as ``Field plane'' in the main text.
The system is designed in a modular fashion, consisting of three main sub-units: endoscope body, calibration module, and laser module.
The output from the laser source is coupled into two polarization-maintaining fibres (Thorlabs PM460-HP, \num{0.12}~NA) which guide a ``signal'' and ``reference'' beams from the laser module to the endoscope body and to the calibration module, respectively.
The calibration module is used only for measuring the transmission matrix of the illumination fibre, being removed from the system once this calibration step preceding imaging is completed.

\subsection*{Multimode fibre}
The illumination multimode fibre (Thorlabs FG200UEA) has a step-index refractive index profile comprising a pure silica core \SI{200}{\micro\metre} in diameter, surrounded by a fluorine-doped silica cladding with \SI{220}{\micro\metre} outer diameter.
The refractive index contrast between the core and cladding materials yields a nominal NA of \num{0.22}, and therefore this fibre sustains approximately \num{17000} guided modes per orthogonal polarization state at the \SI{532}{\nano\metre} wavelength.
The length used was approximately \SI{18.5}{\centi\metre}.
An identical multimode fibre with length of approximately \SI{40}{\centi\metre} was used for collecting the back-scattered signals, and placed adjacent to the illumination fibre (centre-to-centre distance $\approx\SI{250}{\micro\metre}$).

\subsection*{Endoscope body}
In the endoscope body, the output from the ``signal'' polarization-maintaining fibre is collimated by achromatic doublet L1 (Thorlabs AC254-060-A, \(f = \SI{60}{\milli\metre}\)) and illuminates the DMD (ViALUX V-7001) under an incidence angle of \SI{24}{\degree}.
Half-wave plate HW1 (Thorlabs WPMH10M-532) adjusts the angle of polarization incident onto the DMD to optimize the polarization uniformity of the modulated beam reflected off the DMD.
Linear polarizer LP1 (Thorlabs LPVISA100) ensures polarization uniformity and quarter-wave plate QW1 (Newport 10RP04-16) converts the polarization state to circular, which has been shown to be well preserved in step-index multimode fibres \cite{Ploschner2015a}.
Achromatic doublet L2 (Edmund Optics 45-214, $f = \SI{225}{\milli\metre}$) and microscope objective MO1 (Olympus PLN40X, $f = \SI{4.5}{\milli\metre}$) form a 4$f$ system relaying the DMD holograms to the proximal facet of the illumination fibre.
Because the basis of input fields used consisting of truncated plane waves is not strictly orthogonal, best beam-shaping quality is achieved when the number of waveguide modes is slightly oversampled \cite{Turtaev2017}.
For this reason, the particular choice of L2 and MO1 ensures that approximately \num{24000} input modes fit the proximal pupil of the fibre.

In the alternative ``Fourier plane'' configuration, an additional achromatic doublet L2b (Thorlabs AC254-050-A, $f = \SI{50}{\milli\metre}$) is used in combination with L2 to image the holograms onto the back-focal plane of MO1.
In this case, the combination of L2, L2b and MO1 guarantees that around \num{30000} input fields (focal points, in this geometry) fit the fibre core at the proximal facet.

The collection fibre guides the captured back-scattered photons to a GaAsP photomultiplier tube (PMT, Thorlabs PMT2101) equipped with a narrow bandpass filter centred at \SI{531}{\nano\metre} (part of Thorlabs MDF-TOM filter set).
The voltage output of PMT is sampled approximately \num{30} times per DMD hologram by a data acquisition card (National Instruments USB-6351), and the average value is assigned as the grey scale level for that particular pixel in the reconstructed images.

\subsection*{Calibration module}
In the calibration module -- i.e. on the distal side of the endoscope -- microscope objective MO2 (Olympus PLN20X, $f = \SI{9}{\milli\metre}$) together with achromatic doublets L3 (Thorlabs AC254-100-A, $f = \SI{100}{\milli\metre}$) and L5 (Thorlabs AC254-040-A, $f = \SI{40}{\milli\metre}$) image the far field of the illumination fibre onto a CMOS camera (CAM, Basler acA640-750um).
The output from the ``reference'' polarization-maintaining fibre is imaged inside beamsplitter cube BS (Thorlabs BS010) by achromatic doublet L4 (Thorlabs AC254-045-A, $f = \SI{45}{\milli\metre}$) and collimated by L5.
Quarter-wave plate QW2 (Newport 10RP04-16) converts the circularly polarized light emerging from the fibre to linearly polarized.
Half-wave plate HW2 (Thorlabs WPMH10M-532) rotates the angle of polarization of the phase reference beam, and linear polarizer LP2 (Thorlabs LPVISA100) improves the degree of linear polarization, to maximise interference contrast at CAM. 
The combination of MO2, L3 and L5 ensures that the distal pupil of the fibre corresponds to approximately \num{100000} camera pixels, forming an oversampled basis of output fields such that reconstructed images are limited by diffraction rather than by spatial sampling.

All distal optical components and instruments are mounted on the same 3-axis linear translation stage, which conveniently allows removing the calibration module once the transmission matrix of the illumination fibre is measured.

\subsection*{Laser module}
In the laser module, the linearly polarized light output from a single frequency diode-pumped laser source (CrystaLaser CL532-075-S) emitting at the \SI{532}{\nano\metre} wavelength is separated into a signal and reference beams by polarizing beamsplitter PBS (Thorlabs FPB529-23).
Half-wave plate HW3 (Thorlabs WPMH10M-532) placed before Faraday optical isolator ISO (Thorlabs IO-3-532-LP) allows controlling the total optical power in the system, whereas half-wave plate HW4 (Thorlabs WPMH10M-532) located before PBS allows adjusting the relative power between the signal and reference beams.
Achromatic doublets L6 (Thorlabs AC127-025-A-ML, $f=\SI{25}{\milli\metre}$) and L7 (Thorlabs AC254-100-A-ML, $f=\SI{100}{\milli\metre}$) form a magnifying telescope to expand the beam before coupling into the polarization-maintaining fibres by aspheric lenses L8 and L9 (Thorlabs A240TM-A, $f=\SI{8}{\milli\metre}$).
In addition to carrying the ``signal'' and ``reference'' signals to the other system sub-units, the polarization-maintaining fibres also serve as Gaussian spatial filters.
Half-wave plates HW5 and HW6 (Thorlabs WPMH10M-532) are used for precisely aligning the polarization of the coupled beams with the slow axes of the two polarization-maintaining fibres.

\section{\normalsize Holographic methods}
Phase-only modulation is achieved in the off-axis regime from binary-amplitude patterns using the Lee hologram approach \cite{Turtaev2017,Lee1979}, whereby a carrier frequency is added to the desired modulations and only the resulting first diffraction order of the holograms is coupled into the illumination fibre, with the remaining being spatially filtered out by aperture APT (an iris diaphragm) placed in the Fourier plane of L2.
The active area of the DMD was restricted to $768 \times 768$ pixel elements.

Light propagation across the illumination multimode fibre is characterized in terms of its optical transmission matrix \cite{Popoff2010}, acquired empirically by means of phase-step interferometry using an external phase reference.
This arrangement ensures better uniformity in the foci quality across the field of view compared to typical internal phase reference approaches, which give rise to ``blind spots'' in the resulting imagery \cite{Cizmar2011}.
The transmission matrix was measured in the representation of an input basis comprising full DMD gratings of varying orientation and spatial frequency \cite{Turtaev2017}.
In the ``Field plane'' configuration, this corresponds to quasi-orthogonal truncated plane-waves at the input fibre facet, whereas in the conventional ``Fourier plane'' geometry it corresponds to a set of focal points across the proximal fibre endface.
In both cases, the input basis was limited to approximately \num{26000} binary DMD gratings.
The output basis consists of a square grid of far-field foci conjugate to individual camera pixels.
In this convenient representation, the complex modulations for generating each distal far-field focus correspond to individual rows of the transmission matrix acquired.
The calculation of the binary-amplitude DMD patterns further involves an inverse 2D Fourier transform of the complex modulations, and binarizing the resulting phase information \cite{Turtaev2017}.

The interferometric measurements take around \SI{140}{\second} to obtain, limited by the frame rate of the CMOS camera used, and the calculation of all \num{100000} binary patterns takes approximately \SI{150}{\second}.
The entire calibration procedure, including the interferometric measurements, calculation of the transmission matrix, generation of binary patterns, and transferring them to the on-board memory of the DMD, is completed in less than \SI{6}{\minute}.

\section*{\Large Supplementary Results}
\setcounter{section}{0}

\section{\normalsize Imaging at a varying distance}
\label{resolutionmap}

\begin{figure*}[t] 
	\centering
	\captionsetup{justification=justified}
	\includegraphics[width=\textwidth]{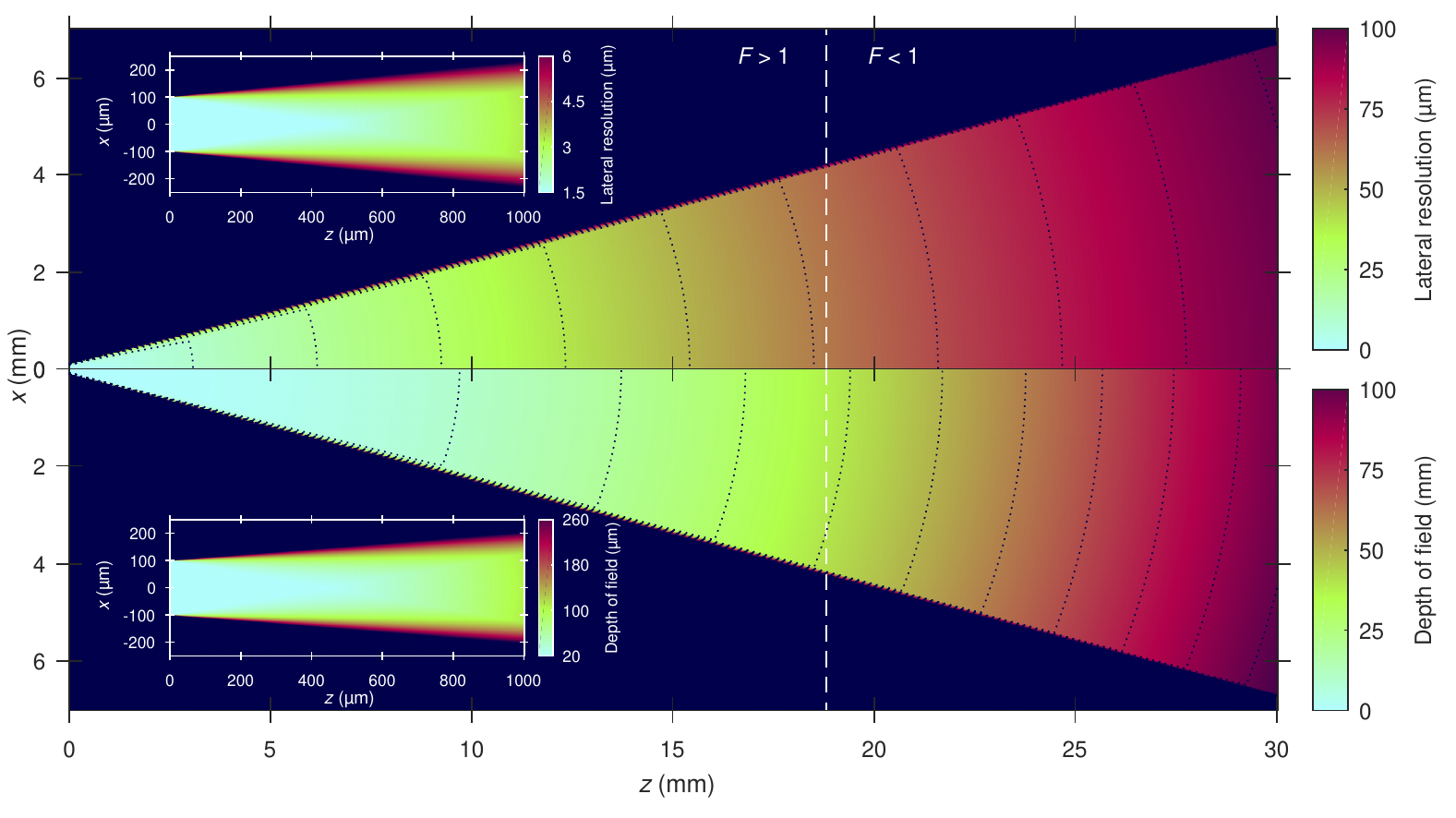}
	\caption{\label{fig:transition}\textbf{Attainable lateral resolution and depth of field.}
		Ray-optics simulation of the lateral resolution and depth of field as function of the imaging distance ($z$) and radial distance from the optical axis ($x$), for a single-fibre endoscope with the same parameters as used in this work (core radius $a=\SI{100}{\micro\metre}$ and $\textrm{NA}=0.22$).
		The dashed vertical line indicates the distance at which the Fresnel number $F$ is 1, separating the region where the far-field condition $F<1$ is fulfilled.		
		The insets show in greater detail the lateral resolution and depth of field attainable in a closer proximity to the distal fibre facet.}
\end{figure*}

Imaging via a multimode fibre holographic endoscope can be performed at any sample plane chosen in front of its distal endface.
The image plane is typically set during the calibration step, with the transmission matrix of the endoscope being measured on a distal plane located at some distance from the output fibre facet.
When this distance is small compared to its core size, output fields carrying all spatial frequencies supported by the fibre are able to interfere constructively across the field of view, and the endoscope can offer diffraction-limited spatial resolution as dictated by the numerical aperture (NA) of the fibre.
With increasing imaging distance, the output fields emerging from the fibre can access a larger field of view, but they are no longer all able to reach the same locations in the sample plane and, as a consequence, the spatial resolution decreases.
This is illustrated by the ray-optics simulation shown in Fig.~\ref{fig:transition}.
For simplicity, we consider here a single fibre endoscope, with the delivery of illumination foci and collection of back-scattered light performed by the same multimode optical fibre.
The collection efficiency of a point source is determined by the intersection of the acceptance cone of the fibre with the spherical sector by which the fibre core is seen from that point.
This allows estimating the effective NA at any point $\left(x,z\right)$ in front of the endoscope.
The lateral resolution is estimated as $\nicefrac{0.61\lambda}{\textrm{NA}}$, and the depth of field (or axial resolution) is given as $\nicefrac{2\lambda}{\textrm{NA}^2}$, where $\lambda$ is the working wavelength. 
In this simulation the fibre has the same parameters as the illumination fibre used in this work (i.e. 0.22~NA and core diameter of \SI{200}{\micro\metre}), the working wavelength is \SI{532}{\nano\metre}, and the surrounding medium has a uniform refractive index of 1 (air).
The dotted curves show equally spaced isolines of both lateral resolution and depth of field.
The vertical dashed line indicates the axial distance $z$ at which the Fresnel number $F$ is 1, a criterion frequently adopted as the far field condition.
The insets in Fig.~\ref{fig:transition} show in greater detail the available resolution and depth of field in close proximity to the endoscope.
The lateral and axial resolutions are uniform and reach their highest values (\SI{1.5}{\micro\metre} and \SI{22}{\micro\metre}, respectively) in a conical region with base corresponding to the circular fibre core, and apex on the optical axis at a distance of $\approx\SI{440}{\micro\metre}$ from the distal fibre facet.

The further the sample plane is located away from the endoscope, the larger becomes the accessible field of view.
On the other hand, the lateral resolution decreases accordingly, since the amount of information transmitted -- closely linked to the number of transmission channels (i.e. waveguide modes) supported by the fibre -- remains the same.
In the far-field imaging regime, the field of view scales linearly (and the lateral resolution inversely) with the imaging distance, which can be seen as a result of the fixed angular resolution of the endoscope.

In the far-field regime, an object can be imaged at any axial distance in the region ($F\lesssim 1$) without modifying the position of the sample plane, due to the high depth of field of the endoscope.
To observe the object at a small imaging distance (i.e. in the region $F\gg1$) where higher spatial resolutions are available, the sample plane needs to be relocated.
This can be achieved with several calibrations of the endoscope (i.e. multiple measurements of the transmission matrix) at different axial planes, or numerically by multiplying the transmission matrix with the appropriate free-space propagation operator \cite{Ploschner2015a}.

\section{\normalsize Tailoring the endoscope performance}
\label{tuning}
\begin{figure*}[h] 
	\centering
	\captionsetup{justification=justified}
	\includegraphics[width=\textwidth]{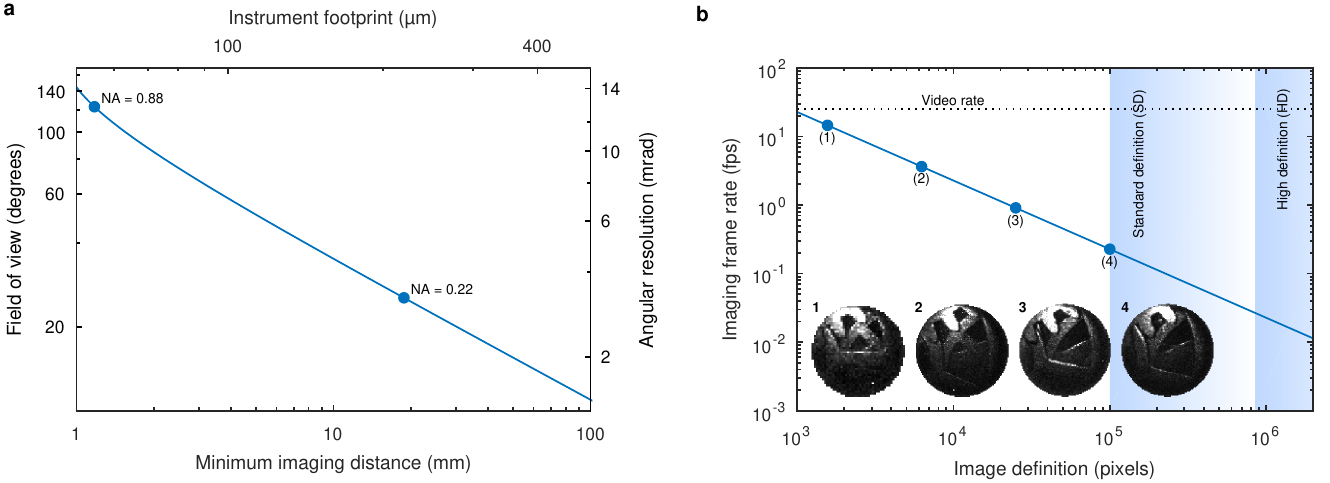}
	\caption{\label{fig:tuning}\textbf{Tailoring the endoscope performance.}
		{\textbf a,} Trade-off between minimum imaging distance, field of view and angular resolution, for different step-index fibres sustaining approximately \num{34000} modes at the \SI{532}{\nano\metre} wavelength.
		The points along the curve represent different combinations of numerical aperture (NA) and core size of the fibres.
		{\textbf b,} Compromise between image definition (i.e. number of pixels) and imaging frame rate.
		The line corresponds to the refresh rate of the DMD employed, \SI{22.7}{\kilo\hertz}.
		Insets (1)-(4) show examples of images obtained using the corresponding sampling regimes marked on the line.}
\end{figure*}

In this work we approach the current technological limits in terms of the number of spatial degrees of freedom controlled -- \num{17000} waveguide modes (per polarization state) sustained by the illumination fibre employed.
Nevertheless, different combinations of fibre parameters impact the imaging properties of the endoscope, which can thus be tailored to fit the needs of particular applications by selecting the suitable multimode fibre.

At a given working wavelength $\lambda$, a step-index fibre with numerical aperture NA and core radius $a$ sustains approximately $\nicefrac{v^2}{2}$ modes, where $v=\left(\nicefrac{2\pi}{\lambda}\right)\cdot a \cdot\textrm{NA}$ is the normalized spatial frequency (or $v$-number) of the waveguide.
In the far-field imaging regime, the angular resolution is dictated by the core size of the illumination fibre, whereas the angular field of view is dictated by its numerical aperture.
The number of resolvable image features thus scales with $\left(a\cdot\textrm{NA}\right)^2$ or, in other words, is proportional to the number of guided modes supported by the fibre, which are indeed the physical channels that transmit image information.
This means that two different illumination fibres supporting the same number of modes will produce identical images -- i.e. with the same definition -- but at different imaging distances.
The plot in Fig.~\ref{fig:tuning}a shows how the field of view, angular resolution, and minimum imaging distance change with the numerical aperture and core size of step-index multimode fibres sustaining approximately \num{34000} guided modes, at the \SI{532}{\nano\metre} wavelength.
The angular field of view is calculated from the numerical aperture of the fibre as $2\sin^{-1}\textrm{NA}$.
Following Rayleigh's criterion, the attainable resolution is estimated as the radius of the Airy disk (i.e. distance from the centre to the first zero-ring of the Airy pattern) forming in the far field, given by $0.61\nicefrac{\lambda}{a}$.
The far field condition can be understood in terms of the Fresnel number, $F$, as $F=\nicefrac{a^2}{d\lambda}\lesssim 1$, where $d$ is the distance to the object.
As such, the smallest imaging distance fulfilling the far field condition is given by $\approx\nicefrac{a^2}{\lambda}$.
The instrument footprint is defined by the cladding diameter of the fibres, which are assumed here as having a 1:1.1 proportion in core-to-cladding diameter (as the fibre used in this work).

The illumination fibre used in this work had a nominal numerical aperture of \num{0.22} and core size of \SI{200}{\micro\metre} in diameter, to which corresponds a field of view of \SI{25}{\degree}, angular resolution of \SI{3.2}{\milli\radian}, and minimum imaging distance of \SI{19}{\milli\metre}, as highlighted in the plot of Fig.~\ref{fig:tuning}a.
A high-NA fibre with \num{0.88}~NA and \SI{50}{\micro\metre} core diameter, also highlighted in Fig.~\ref{fig:tuning}a, sustains an identical number of guided modes at the same working wavelength, thus yielding images with the same definition.
In this case, the field of view would be \SI{123}{\degree}, the angular resolution \SI{13}{\milli\radian}, the minimum imaging distance \SI{1.2}{\milli\metre}, and the instrument footprint area reduced by 16 fold.

The compromise between image definition (i.e. number of image pixels, $N$) and image acquisition frame rate is illustrated in Fig.~\ref{fig:tuning}b.
When employing a spatial light modulator with refresh rate $R$, the acquisition frame rate of full images is given simply by $\nicefrac{R}{N}$.
The solid line in Fig.~\ref{fig:tuning}b shows the possible combinations between image definition and their acquisition frame rate when operating the DMD modulator at its maximum refresh rate of \SI{22.7}{\kilo\hertz}.
The points highlighted on this line correspond to the different spatial sampling regimes shown in the respective figure insets as well as in Supplementary Movie SM2.
Random access to any positions in the hologram sequence uploaded to the DMD on-board memory allows restricting scanning to any number of image pixels (e.g. in a particular region of interest within the field of view), making it possible to further increase the acquisition frame rate beyond these values.
The 16~GB on-board memory of the DMD modulator allows storing \num{174762} binary patterns with size 1024$\times$768, which represents the current practical limit for the image definition of the endoscope.

\section{\normalsize Additional considerations on power ratio}
\label{powerratio}
We start by considering that, at any given plane perpendicular to the optical axis (oriented along the $z$ direction), the target field $\left\vert E_T\right\rangle = E_T\left(x,y\right)$ which one wishes to synthesize differs from the actual modulated field $\left\vert E_M\right\rangle = E_M\left(x,y\right)$ that is obtained by wavefront shaping.
We can express the modulated field $\left\vert E_M\right\rangle$ as:
\begin{equation}
\left\vert E_M\right\rangle = \alpha \left\vert E_T\right\rangle + \left(\left\vert E_M\right\rangle - \alpha \left\vert E_T\right\rangle\right),
\label{eq:Emodulated}
\end{equation}
and set the value of $\alpha$ in such way that the two terms on the right-hand side of Eq.~\eqref{eq:Emodulated} are orthogonal, which yields $\alpha = \nicefrac{\left\langle E_T \vert E_M \right\rangle}{\left\langle E_T \vert E_T \right\rangle}$.
Now we can re-write Eq.~\eqref{eq:Emodulated} as:
\begin{equation}
\left\vert E_M\right\rangle = \left\vert E_{\textrm{in}} \right\rangle + \left\vert E_{\textrm{out}} \right\rangle,
\end{equation}
where:
\begin{align}
\left\vert E_{\textrm{in}} \right\rangle &= \frac{\left\langle E_T \vert E_M \right\rangle}{\left\langle E_T \vert E_T \right\rangle} \left\vert E_T \right\rangle, \\
\left\vert E_{\textrm{out}} \right\rangle &= \left\vert E_M \right\rangle - \frac{\left\langle E_T \vert E_M \right\rangle}{\left\langle E_T \vert E_T \right\rangle} \left\vert E_T \right\rangle,
\end{align}
and with $\left\langle E_{\textrm{in}} \vert E_{\textrm{out}} \right\rangle = 0$.
The field $\left\vert E_{\textrm{in}} \right\rangle$ (the projection of $\left\vert E_M \right\rangle$ onto the target field $\left\vert E_T \right\rangle$) represents the portion of the actual field $\left\vert E_M \right\rangle$ which contributes to the desired field $\left\vert E_T \right\rangle$.
The field $\left\vert E_{\textrm{out}} \right\rangle$, on the other hand, being orthogonal to $\left\vert E_{\textrm{in}} \right\rangle$, represents the amount of the field $\left\vert E_M \right\rangle$ which does not contribute to the target field $\left\vert E_T \right\rangle$.

The power ratio, $\eta$, is defined as the fraction of total power which contributes to the target field.
In other words, it is given by the ratio between the power contained in the field $\left\vert E_{\textrm{in}} \right\rangle$, $P_{\textrm{in}} = \left\langle E_{\textrm{in}} \vert E_{\textrm{in}}\right\rangle$, and the total power contained in the actual modulated field $\left\vert E_M \right\rangle$, $P_{\textrm{tot}} = \left\langle E_M \vert E_M\right\rangle$.
Thus:
\begin{equation}
\eta = \frac{P_{\textrm{in}}}{P_{\textrm{tot}}}
= \frac{\left\langle E_{\textrm{in}} \vert E_{\textrm{in}}\right\rangle}{\left\langle E_M \vert E_M\right\rangle}
= \frac{\left\langle E_T \vert E_M\right\rangle^{*} \left\langle E_T \vert E_M\right\rangle}{\left\langle E_T \vert E_T\right\rangle \left\langle E_M \vert E_M\right\rangle}.
\label{eq:powerRatio}
\end{equation}

If $\left\vert \Psi_T \right\rangle$ and $\left\vert \Psi_M \right\rangle$ are the normalized fields $\left\vert E_T \right\rangle$ and $\left\vert E_M \right\rangle$, then the power ratio can be written simply as:
\begin{equation}
\eta = \left\vert \left\langle \Psi_T \vert \Psi_M \right\rangle \right\vert^2.
\end{equation}
This shows that the power ratio $\eta$ can be understood simply as the squared modulus of the overlap integral between the ideal target field $\left\vert \Psi_T \right\rangle$ and the actual synthesized field $\left\vert \Psi_M \right\rangle$, which is a measure of their similarity.

\bibliographystyle{naturemag_noURL}
\bibliography{PepperRefs.bib}

\clearpage
\onecolumn
\section*{\Large Supplementary Media}
\renewcommand\thefigure{SM\arabic{figure}}
\renewcommand{\figurename}{Supplementary Movie}
\setcounter{figure}{0}

\begin{figure*}[h] 
	\centering
	\captionsetup{justification=justified}
	\includegraphics[width=0.4\textwidth]{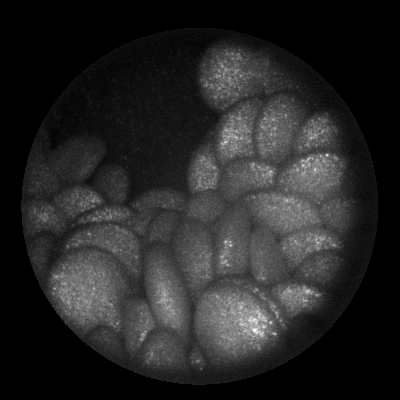}
	\caption{\label{fig:pepper}\textbf{Multimode fibre based endoscopic imaging inside a sweet pepper.}
		The video recording shows macroscopic imaging of an approaching sweet pepper serving the purpose of a medical imaging phantom.
		A small opening on its side facilitates access of the endoscope to its interior, where multiple seeds are observed in detail.}
\end{figure*}

\begin{figure*}[h] 
	\centering
	\captionsetup{justification=justified}
	\includegraphics[width=0.4\textwidth]{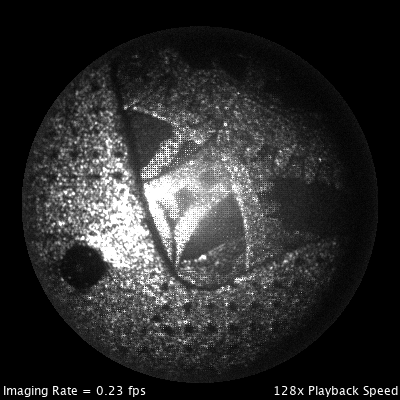}
	\caption{\label{fig:clock}\textbf{Multimode fibre based endoscopic imaging of a functioning clockwork mechanism.}
		The video recording shows macroscopic imaging of a functioning mechanical clock, in particular of its gear mechanism at work.
		Motion artefacts become apparent in the parts displaying faster dynamics.
		To mitigate this effect, the spatial sampling is decreased on-the-fly to increase the image acquisition frame rate, better revealing the dynamics of the object down to the \SI{100}{\milli\second} time scale.
	}
\end{figure*}